\def\pnn{K_L \rightarrow \pi^0\nu\overline{\nu
}}
\def\ppX{K_L \rightarrow \pi^0 \pi^0 X}
\def\ppp{K_L \rightarrow 3\pi^0 }

\def\ppgg{K_L \rightarrow \pi^0 \pi^0 \gamma \gamma }

\documentclass[twocolumn, superscriptaddress, aps,prl]{revtex4-2}
\newcommand{\mycomment}[1]{}

\usepackage{graphicx}% Include figure files
\usepackage{dcolumn}% Align table columns on decimal point
\usepackage{bm}% bold math
\usepackage{amsmath}
\usepackage{xcolor}
\usepackage{booktabs}
\usepackage{siunitx}
\usepackage[caption=false]{subfig} % for \subfloat
\usepackage{setspace}
\begin{document}
\setlength{\columnsep}{25pt}
%\linenumbers
\preprint{APS/123-QED}
\title{The Search for $K_L \rightarrow \pi^0\pi^0\gamma\gamma$ and $K_L\rightarrow \pi^0\pi^0X$ where $X\rightarrow 2\gamma$ at the KOTO Experiment}

\author{J.~Redeker}
\affiliation{Enrico Fermi Institute, University of Chicago, Chicago, Illinois 60637, USA}
\author{C.~Lin}
\affiliation{Department of Physics, National Changhua University of Education, Changhua 50007, Taiwan}
\author{Y.~W.~Wah}
\affiliation{Enrico Fermi Institute, University of Chicago, Chicago, Illinois 60637, USA}
\author{J.~K.~Ahn}
\affiliation{Department of Physics, Korea University, Seoul 02841, Republic of Korea}
\author{M.~Gonzalez}
\altaffiliation[]{Present address: CERN, European Organization for Nuclear Research, CH-1211 Geneva, Switzerland.}
\affiliation{Department of Physics, Osaka University, Toyonaka, Osaka 560-0043, Japan}
\author{K.~Hanai}
\affiliation{Department of Physics, Osaka University, Toyonaka, Osaka 560-0043, Japan}
\author{Y.~B.~Hsiung}
\affiliation{Department of Physics, National Taiwan University, Taipei 10617, Taiwan, Republic of China}
\author{T.~Kato}
\affiliation{Department of Physics, Osaka University, Toyonaka, Osaka 560-0043, Japan}
\author{E.~J.~Kim}
\affiliation{Division of Science Education, Jeonbuk National University, Jeonju 54896, Republic of Korea}
\author{T.~K.~Komatsubara}
\affiliation{Institute of Particle and Nuclear Studies, High Energy Accelerator Research Organization (KEK), Tsukuba, Ibaraki 305-0801, Japan}
\affiliation{J-PARC Center, Tokai, Ibaraki 319-1195, Japan}
\author{K.~Kotera}
\affiliation{Department of Physics, Osaka University, Toyonaka, Osaka 560-0043, Japan}
\author{S.~K.~Lee}
\affiliation{Division of Science Education, Jeonbuk National University, Jeonju 54896, Republic of Korea}
\author{G.~Y.~Lim}
\affiliation{Institute of Particle and Nuclear Studies, High Energy Accelerator Research Organization (KEK), Tsukuba, Ibaraki 305-0801, Japan}
\affiliation{J-PARC Center, Tokai, Ibaraki 319-1195, Japan}
\author{T.~Matsumura}
\affiliation{Department of Applied Physics, National Defense Academy, Kanagawa 239-8686, Japan}
\author{H.~Nanjo}
\affiliation{Department of Physics, Osaka University, Toyonaka, Osaka 560-0043, Japan}
\author{T.~Nomura}
\affiliation{Institute of Particle and Nuclear Studies, High Energy Accelerator Research Organization (KEK), Tsukuba, Ibaraki 305-0801, Japan}
\affiliation{J-PARC Center, Tokai, Ibaraki 319-1195, Japan}
\author{T.~Nunes}
\affiliation{Department of Physics, Osaka University, Toyonaka, Osaka 560-0043, Japan}
\author{K.~Ono}
\affiliation{Department of Physics, Osaka University, Toyonaka, Osaka 560-0043, Japan}
\author{K.~Shiomi}
\affiliation{Institute of Particle and Nuclear Studies, High Energy Accelerator Research Organization (KEK), Tsukuba, Ibaraki 305-0801, Japan}
\affiliation{J-PARC Center, Tokai, Ibaraki 319-1195, Japan}
\author{R.~Shiraishi}
\affiliation{Institute of Particle and Nuclear Studies, High Energy Accelerator Research Organization (KEK), Tsukuba, Ibaraki 305-0801, Japan}
\author{Y.~Tajima}
\affiliation{Department of Physics, Yamagata University, Yamagata 990-8560, Japan}
\author{Y.~C.~Tung}
\affiliation{Department of Physics, National Kaohsiung Normal University, Kaohsiung 824, Taiwan}
\author{H.~Watanabe}
\affiliation{Institute of Particle and Nuclear Studies, High Energy Accelerator Research Organization (KEK), Tsukuba, Ibaraki 305-0801, Japan}
\affiliation{J-PARC Center, Tokai, Ibaraki 319-1195, Japan}
\author{T.~Wu}
\affiliation{Department of Physics, National Taiwan University, Taipei 10617, Taiwan, Republic of China}
\author{T.~Yamanaka}
\affiliation{Department of Physics, Osaka University, Toyonaka, Osaka 560-0043, Japan}
\author{H.~Y.~Yoshida}
\affiliation{Department of Physics, Yamagata University, Yamagata 990-8560, Japan}

%\affiliation{
%\affiliation{
%\affiliation{%


\begin{abstract}
 We performed searches for $\ppX$ where $X$ may be an axion-like particle which promptly decays to two photons, and the first search for $\ppgg$ at the KOTO experiment using data taken in 2021. The search is performed for $X$ mass in the range of 160--220 MeV/$c^2$. Three events were observed in the signal region, with two events near an $X$ mass of 177 MeV/$c^2$. This result led to a range of upper limits on the branching ratio, BR($\ppX$) $< (1\text{--}20) \times 10^{-7}$ at the 95\% confidence level (C.L.). No events were observed for the analysis of $\ppgg$, setting an upper limit on the branching ratio, BR($\ppgg$) $< 1.69 \times 10^{-6}$ at the 95\% C.L.  

\end{abstract}

\maketitle

%\tableofcontents

%\section{\label{sec:level1}First-level heading:\protect\\ The line
%break was forced \lowercase{via} \textbackslash\textbackslash}
%\section{\label{sec:intro} \textit{Introduction--}}
\textit{Introduction---} Searching for dark sector particles is one of the major tasks in particle physics today. There is theoretical interest in dark particle candidates that couple to quarks in an energy range that a neutral-kaon experiment is sensitive to through the study of rare decays~\cite{qcd_scale_alp_coupling, dark_matter_quark_coupling, dark_matter_chiral, Bauer_2022}. A promising search is the decay $\ppX$, where $X$ may be an axion-like particle (ALP) which decays promptly. In particular, there are theoretical scenarios that prefer three-body decay modes such as $\ppX$ over $K_L\rightarrow\pi^0X$~\cite{flavourviolation_alp_balkin, balkin_pion_chimney}. This decay may be observable if $X$ decays to two photons via a quark loop, and if X has a mass different from the $\pi^0$. $\ppX$ was studied in the E391a experiment \cite{yuchen} in a smaller mass range, $M_X \in [194.3,219.3]$ MeV/$c^2$. This study improves sensitivity up to an order of magnitude in the mass range studied in E391a, while also providing a first search for $M_X$ closer to the $\pi^0$ mass. This analysis also provides insight into the rare decay $\ppgg$. This decay is described by Chiral Perturbation Theory (ChPT) \cite{funck_kambor} up to second order, and a measurement of the branching ratio provides a test of second order ChPT. Additionally, this decay is closely related to the theoretical branching ratio for modes $K_L\rightarrow\pi^0\pi^0l^-l^+$ which were studied in the KTeV experiment~\cite{2pi0ee_KTeV, 2pi0mumu_KTeV}, where the calculation of the branching ratio is fundamentally affected by the process $K_L\rightarrow\pi^0\pi^0\gamma^*\gamma^*$, where $\gamma^*$ is a virtual photon.   

\begin{figure*}
\includegraphics[width=1\linewidth]{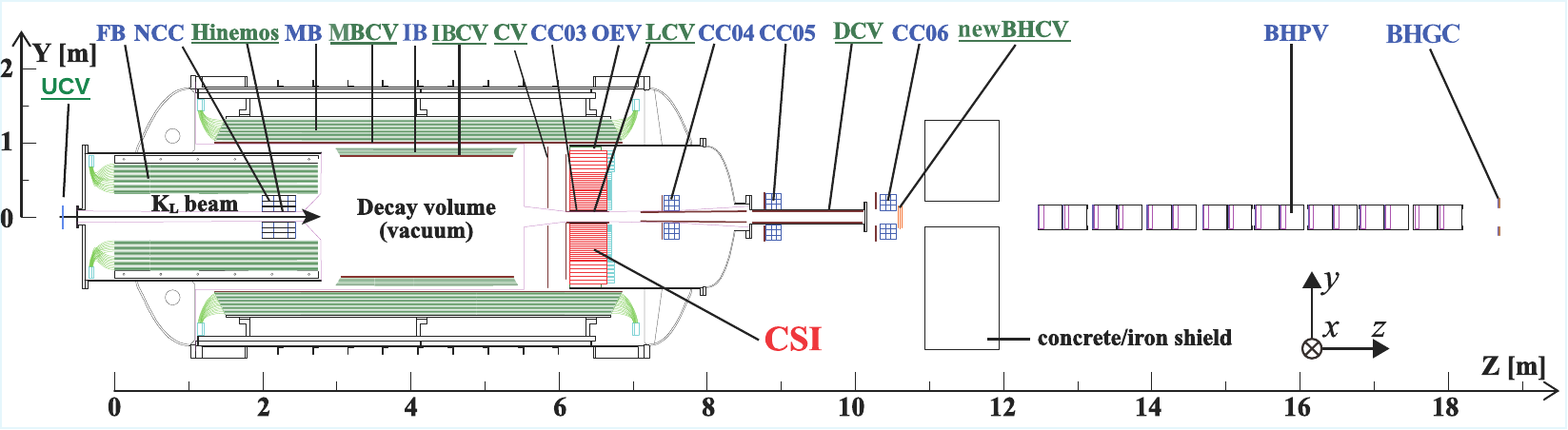}
  \caption{Cross-sectional side view of KOTO detector. A detector name with an underline implies it is a charged particle veto counter. All others, except for the CSI calorimeter, are photon veto counters.} 
  \label{fig:KOTODetector}
\end{figure*}

The primary purpose of the KOTO experiment at the Japan Proton Accelerator Research Complex (J-PARC) facility in Tokai Japan is to search for the ultra-rare decay $\pnn$ \cite{Buras, pi0nunu_2021}. KOTO uses the intense 30-GeV proton beam which is extracted from the main ring towards a gold target in the Hadron Experimental Facility that hosts the KOTO detector. The secondary 20-meter neutral beamline is constructed at an angle of 16$^\circ$ off the proton beam-axis and collimated with a solid angle of \SI{7.8}{\micro\steradian} which corresponds to an $8\times8\text{ cm}^2$ beam width at the exit of the second collimator \cite{shimogawa_beamline_design}. A photon absorber made of 70--mm thick lead was inserted in front of the collimators to reduce the number of photons entering the detector volume. A sweeping magnet was installed between the two collimators to reduce the number of charged particles entering the detector volume. The $K_L$ momentum spectrum peaks at 1.4 GeV/$c$ at the entrance of the detector and was measured during an engineering run in 2012 using the $K_L \rightarrow\pi^+\pi^-\pi^0$ decay \cite{sato_thesis, kl_flux_momentum}. A cross-sectional schematic of the KOTO detector is shown in Fig.~\ref{fig:KOTODetector}. 
The origin is set on the beam axis and at the upstream side of the Front Barrel (FB). The calorimeter (CSI) \cite{csi_calorimeter} has a diameter of \SI{1.9}{\meter} and a depth of 27 radiation lengths. The primary purpose of the calorimeter is to detect photons from $K_L$ decays, and measure  their position, energy, and timing. The calorimeter consists of an array of 2716 undoped CsI crystals with the dimensions of $2.5\times2.5\times50$ cm$^3$ (inner) and $5\times5\times50$ cm$^3$ (outer). KOTO uses a hermetic veto system to detect all particles which arise from a $K_L$ decay within the detector region. The largest veto detectors consist of lead plastic-scintillator sandwich counters that enclose the decay volume (IB, MB, and FB) \cite{barrel_veto, barrel_vetoes}. Plastic scintillators to detect charged particles are placed on the inner surface of IB and MB, as well as in front of CSI (IBCV, MBCV, and CV) \cite{cv}. Other veto detectors such as the Neutron Collar Counter (NCC) and Collar Counters CC03--6 are made with additional undoped CsI crystals that are placed near the beam axis. The beam-hole veto detectors consist of three layers of wire chambers for the detection of charged particles (newBHCV) and an arrangement of aerogel Cherenkov counters with lead converters for the detection of photons (BHPV) \cite{bhpv_maeda}. The data acquisition system relies on two stages of trigger logic \cite{cluster_finding_trigger}. The first trigger stage (L1) required the total deposited energy in CSI to be larger than 550 MeV and no hits in NCC, MB, IB, CV, CC03, CC04, CC05, or CC06 detectors. The second trigger stage (L2) selected events based on the number of clusters of deposit energy in CSI. For the study of $\ppgg$ and $\ppX$, the number of clusters was required to be six. The same trigger condition was used to study the $\ppp$ decay to evaluate the background level and the $K_L$ yield. Once a trigger condition was passed, the digitized waveforms of each channel was recorded for offline analysis. 

%\section{Event Reconstruction}

\textit{Event Reconstruction---}Event reconstruction involves calculating complex analysis variables from the measured waveforms in data. For $\ppgg$, there is no missing information since all final state particles are measured. There are 45 combinations ($_6\text{C}_2 \times {}_4\text{C}_2/2$) of two photons to pair with the two pions
(the factor of two comes from the redundancy of pairing). A constrained fit strategy is used for the reconstruction of $\ppgg$. The energy and hit position of the photons in the CSI should be well reconstructed and follow conservation of momentum and energy constraints. In total, there are five constraints applied in the reconstruction. First, the reconstructed invariant mass of the six photons should be the $K_L$ mass. The second and third constraints require the invariant mass of two of the photon pairs to be the pion mass. Lastly, the fourth and fifth constraints require the $K_L$ momentum vector to lie on the line connecting the gold target and the Center of Energy (CoE) on CSI. 
\mycomment{The process for this reconstruction is done in two stages. In the first stage, the reconstructed hit positions and energies of photons in the CSI are calculated, and the pion decay vertex is obtained by assuming the pion mass for each of the two pions as
\begin{equation}
    M_{\pi^0}^2 = 2E_1E_2(1-\cos\theta),
    \label{eq:pion_mass}
\end{equation}
where $E_1$ and $E_2$ represent the energy of the paired photons, and $\theta$ is the opening angle between them. The preliminary $K_L$ decay vertex is the average of reconstructed pion vertices on the beam-axis, and the kinematics of the pions and kaon are reconstructed under this assumption.}
The constraints are used to modify the reconstruction variables, such as the decay vertex, kaon and pion momentum, and photon hit position and energy, to best fit the constraints. Knowing the position and energy resolution in the calorimeter, the ``goodness'' of the constrained fit can be quantified in a variable $\chi^2$. The $\chi^2$ is a function of the photon $x,y$ hit positions on CSI, and the photon energies, and is defined with known resolutions of hit position ($\sigma_{x_i}, \sigma_{y_i}$) and energy ($\sigma_{E_i}$) as
{\small
\begin{align}
    \chi^2_{fit}&(x_1,y_1,E_1,...,x_6,y_6,E_6) = \notag \\ &\sum_{i=1}^6\frac{(x_i - x_{i,m})^2}{\sigma_{x_i}^2} +\sum_{i=1}^6\frac{(y_i - y_{i,m})^2}{\sigma_{y_i}^2} + \sum_{i=1}^6\frac{(E_i - E_{i,m})^2}{\sigma_{E_i}^2} \hspace{0.2cm}. 
    \label{eq:chisq_function}
\end{align}}
The variables $x_{i,m}, y_{i,m}$ and $E_{i,m}$ represent the initial measured reconstructed position and energy of the photon hit, while $x_i, y_i$ and $E_i$ are the fitted results obtained by minimizing the $\chi^2_{fit}$ function and are used in the following analysis. Out of the 45 possible combinations, the pairing which returns the smallest $\chi^2_{fit}$ value was chosen as the best fit for the reconstruction. This $\chi^2_{fit}$ was obtained using the signal constraints, and will be denoted as $\chi^2_{fit}(2\pi^0\gamma\gamma)$. This reconstruction strategy can likewise be done assuming the decay $\ppp$. In this case, there is an additional pion mass constraint, and the photon pairings which give the smallest $\chi^2_{fit}$ with an additional pion mass assumption is recorded ($\chi^2_{fit}(3\pi^0)$). If $\chi^2_{fit}(3\pi^0)$ is small, it is likely a background event from $\ppp$, whereas for the signal this value should be large as long as the $X$ mass is dissimilar from the pion mass.

\begin{figure}[h]
\includegraphics[width=0.43\textwidth]{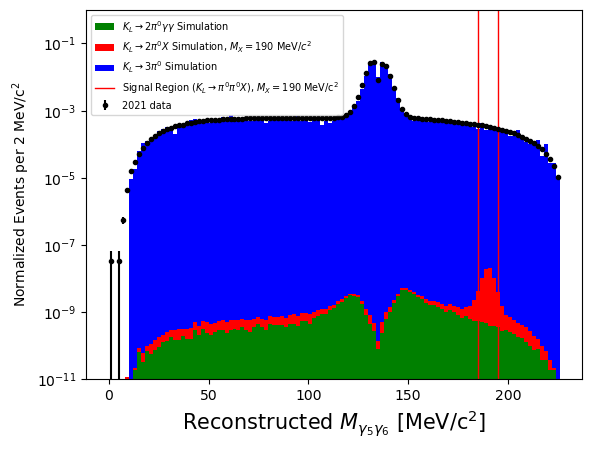}
  \caption{Invariant mass, $M_{\gamma_5\gamma_6}$, distribution in data, overlaid with the $\ppp$ background simulation normalized to the branching ratio. The signal simulation is weighted according to an assumed $\text{BR}(\ppX) = 1 \times 10^{-7}$ and $\text{BR}(\ppgg) = 8.4\times 10^{-8}$ \cite{funck_kambor}.}
  \label{fig:mg5g6_example}
\end{figure}

The simulation of the decay $\ppX$ where $X$ decays to $2\gamma$ promptly was performed assuming a flat phase space. Following the same reconstruction, the signal region is defined by the invariant mass of the two photons not used in the pion reconstruction, called $M_{\gamma_5\gamma_6}$. The data overlaid with the simulated Monte Carlo distributions are shown in Fig.~\ref{fig:mg5g6_example} after applying basic selection criteria (cuts) for veto and data quality.
\mycomment{There are two notable features of the $M_{\gamma_5\gamma_6}$ distribution for $\ppp$. The first is the dip at $M_{\pi^0}$ which arises due to the fact that the two other photon pairings will have invariant masses closer to $M_{\pi^0}$ than the last pairing plotted in Fig.~\ref{fig:mg5g6_example}. The second important feature is the large invariant mass tails. The smaller high mass tail is the motivation to focus on the high mass region for this search. Nonetheless, if the $\ppp$ background is not well controlled, the huge background tail may easily overwhelm the signal.} The simulation of the decay $\ppgg$ was performed following the phase space defined in \cite{funck_kambor}. Notably, the invariant mass distribution has a pole near the pion mass, and so a precut is applied in the simulation around the pion mass, $\left|M_{\gamma_5\gamma_6} - M_{\pi^0}\right| < 10$ MeV/$c^2$.

%\section{Selection Process}

\textit{Selection Process---}General criteria were applied to ensure data quality in this analysis. Six clusters in the CSI with no on-time hits in the veto detectors, shown in Fig.~\ref{fig:KOTODetector}, was required.  The photon hit position $(x,y)$ in CSI were restricted such that, $\max(|x|,|y|) > 150$ mm and $\sqrt{x^2+y^2} < 850$ mm in order to prevent shower leakage on the edge. An event was rejected if the hit position was in the region $(210\text{ mm}< x < 260 \text{ mm }) \wedge (100\text{ mm} < y < 135\text{ mm})$ to avoid CSI regions more likely to induce energy mismeasurement caused by unstable, low gain PMTs. This region was determined by looking at the reconstructed hit positions of $\gamma_5$ and $\gamma_6$ with large $\chi^2_{fit}(3\pi^0)$ values in data. The minimum photon energy was required to be larger than $100$ MeV, and the minimum cluster distance was required to be larger than 100 mm in order to ensure the individual clusters are well measured. The total energy deposit in CSI was required to be larger than 650 MeV. The reconstructed $K_L$ transverse momentum should be less than 8 MeV/$c$ to protect against energy mismeasurement. The photon hit timing must be within 1 ns of each other. The cluster shape and channel pulses were required to be consistent with a library of photon-like cluster shapes and waveforms, with the discriminators called CSDDL and FPSD respectively \cite{csddl_fpsd}. 
%\section{$K_L$ Yield}

\textit{Selection Optimization---} To determine the event selection, ten times the statistics of $\ppp$ in data was simulated. This simulation sample was used to determine the selection criteria for the 13 $\ppX$ analyses with $X$ masses in 5 MeV intervals ranging from 160--220 MeV/$c^2$, and for $\ppgg$. The cut optimization focused on the three most important variables in these analyses, $\chi^2_{fit}(3\pi^0)$, $\chi^2_{fit} (2\pi^0\gamma\gamma)$, and a deep learning cluster shape discriminator (CSDDL) \cite{csddl_fpsd}. CSDDL was originally trained to discriminate between neutron and electromagnetic interactions in CSI. However, this variable is highly effective because the photonuclear interactions in CSI will likewise affect the cluster shape, and energy mismeasurement due to the photonuclear effect is the largest cause of $\ppp$ background in this analysis. A three-dimensional grid search was done to test each set of cuts and evaluate them based on a Figure of Merit (FOM) strategy. A FOM strategy optimizes the potential upper limit (U.L.) result by maximizing a function \cite{FOM_Cowan}
\begin{align}
    f(N_{sig}, N_{bkg}) &= \sqrt{2\ln(Q)}\hspace{0.1cm}, \\\hspace{0.1cm} Q &= e^{-N_{sig}}\left(1 + \frac{N_{sig}}{N_{bkg}}\right)^{N_{sig} + N_{bkg}} \hspace{0.3cm}, 
    \label{eq:fom_function}
\end{align}
where $N_{sig}$ and $N_{bkg}$ were calculated within the region of interest (ROI) of $M_{\gamma_5\gamma_6}$ and represent the number of signal and background events in the ROI, respectively. $N_{sig}$ was calculated assuming a branching ratio of the signal at $10^{-7}$. For the X mass analyses, the ROI was found by applying a Gaussian fit on $M_{\gamma_5\gamma_6}$ to the signal simulation, and was defined by $\mu_{fit} \pm 2\sigma_{fit}$. For the $\ppgg$ analysis, the ROI was defined by the high mass region $M_{\gamma_5\gamma_6} \in[160, 227.66]\text{ MeV/}c^2$. By optimizing the event selection process for each analysis, the signal acceptance can be improved by up to a factor of five for $X$ masses near the 160 MeV/$c^2$ and 220 MeV/$c^2$.

\textit{Control Region Analysis---} In this study, the $M_{\gamma_5\gamma_6}$ region above 155 MeV/$c^2$ was masked to avoid human bias. Because of this, the region with $M_{\gamma_5\gamma_6}$ below 130 MeV/$c^2$ was crucial for understanding the agreement between data and Monte Carlo. If the photonuclear interaction cross section in CSI is larger than predicted by Monte Carlo simulation in KOTO's energy range, we would underestimate $\ppp$ background levels. A scale factor may be determined by studying the low mass region after tight selections in data. This factor was simply defined as $F = N_{\mathrm{obs}}^{\mathrm{data}}/N_{\mathrm{exp}}^{\mathrm{MC}}$, where $N_{\mathrm{obs}}^{\mathrm{data}}$ and $N_{\mathrm{exp}}^{\mathrm{MC}}$ are the number of events observed in data and the number of events expected from Monte Carlo simulation, respectively. Correlations were observed between CSDDL, $\chi^2_{fit}(3\pi^0)$, and $F$. Since each analysis has different selection criteria, the scale factor, $F$, was determined independently for each analysis. This factor was typically in the range of 1.5--2.5. The fidelity of this method to evaluate the background level was tested using an inverse selection strategy in the high mass region.
\begin{figure}[h]
\includegraphics[width=0.4\textwidth]{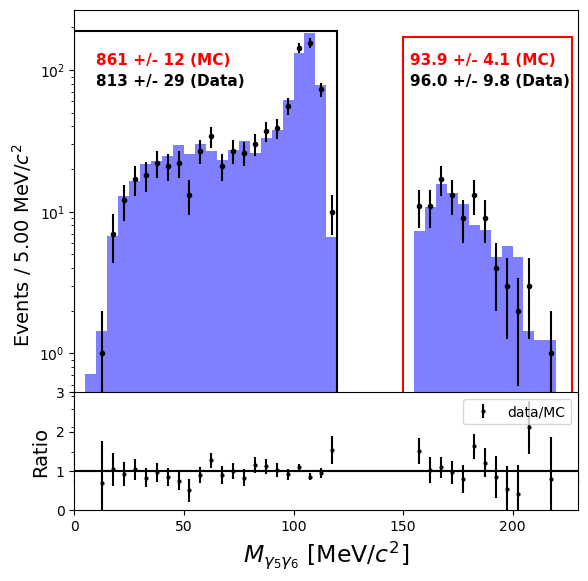}
  \caption{$\ppp$ data (black) and Monte Carlo (blue) comparison after applying the inverse selection and scale factor. The numbers represent the number of data and Monte Carlo events observed in the regions encapsulated by the black (red) lines. }
  \label{fig:inverse_cut}
\end{figure}
Events were required to have $100 < \chi^2_{fit}(3\pi^0) < 115$ to observe the performance at large $\chi^2_{fit}(3\pi^0)$ values while excluding events that may fall in the masked region. All other cut selection criteria are applied except for FPSD. The CSDDL cut was loosened in order to enhance statistics. $F$ was determined using ~50\% of the $\ppp$ Monte Carlo. The agreement between data and Monte Carlo are observed in Fig.~\ref{fig:inverse_cut}, tested against 100\% of the $\ppp$ Monte Carlo. 

\begin{figure}[h]
\includegraphics[width=0.45\textwidth]
{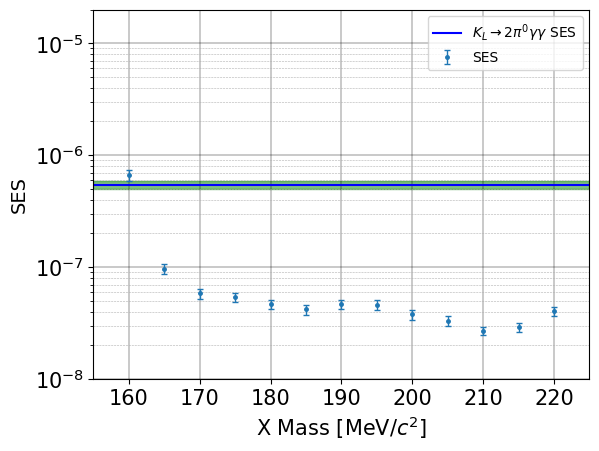}
  \caption{SES as a function of $X$ mass. The horizontal bar represents the SES of the $\ppgg$ decay. The error bars represent the quadratic sum of the statistical and systematic uncertainty which is propagated from the $K_L$ yield and $A_{sig}$.}
  \label{fig:SES}
\end{figure}

\textit{Single Event Sensitivity---}The set of data analyzed in this article was taken in 2021 with a proton beam intensity between 60--64 kW, and corresponds to a total $K_L$ yield at the entrance of the KOTO detector, $Y$, given by  
 \begin{equation}
    Y = (4.27 \pm 0.03_{stat} \pm 0.31_{sys}) \times 10^{12} \hspace{0.2cm}. 
    \label{eq:3pi0_yield_result}
\end{equation}
The yield was determined through the copious $\ppp$ decay, then normalized by calculating its acceptance using Monte Carlo simulation \cite{geant4_simulation_toolkit}. The sensitivity of the analysis result is quantified in the Single Event Sensitivity (SES), which represents the central value of the branching ratio if one signal event is observed, and is defined as 
\begin{equation}
    \mathrm{SES}  = \frac{1}{A_{sig}\times Y} \hspace{0.2cm} \hspace{0.2cm}.
    \label{eq:SES}
\end{equation}
The systematic uncertainty in this calculation is summarized in Table~\ref{tab:sys_error}. 
\begin{table}[h] 
\caption{Systematic uncertainties associated with the SES calculation. }
\begin{center}
 \begin{tabular}{|| l | c ||} 
 \hline 
 \textbf{Source} & Uncertainty\\ %[0.5ex] 
 \hline\hline
 Veto Cuts & 4.6\% \\ 
 \hline
 Kinematic Cuts & 2.3\%   \\
 \hline
 Quality Cuts & 2.9\%  \\
 \hline
  $K_L$ Momentum  & 1.5\%  \\ 
 \hline
  Trigger & 4.1\%   \\ 
  \hline 
  $\chi^2_{fit}(3\pi^0)$ & $5\text{--}8.8 \%$ \\
 \hline
 \hline
 \textbf{Total} & \textbf{8.9--11.5\% } \\
 \hline
\end{tabular}
 \label{tab:sys_error}
\end{center}
\end{table}
The dominant sources of systematic uncertainties are from the veto cut selection (5.1\%) and the trigger effect (4.1\%). The systematic uncertainties of the veto cut selection, the kinematic cut selection (2.3\%), and the $\chi^2$ cut selections (1.7\%), were determined by quadratically summing the relative difference in exclusive acceptance between data and Monte Carlo simulation, which is defined as $A_i = N_{\mathrm{all}}/N_{\mathrm{all \backslash i}}$ where $N_{\mathrm{all}}$ and $N_{\mathrm{all \backslash i}}$ are the number of events observed after applying all event selections and the number of events observed after applying all but the $i$-th, respectively. The systematic uncertainty from the $K_L$ momentum spectrum is related to the geometrical acceptance, which is the probability to observe all final state particles in CSI, and the kinematic acceptance. The simulated $K_L$ momentum spectrum relies on parameters obtained during engineering runs in 2012 \cite{sato_thesis,kl_flux_momentum}. The magnitude of this effect was obtained by calculating the difference in Monte Carlo geometrical acceptance by varying the parameters according to their measured uncertainty. The 1$\sigma$ variation in the relative difference in acceptance was found to be 1.5\%. The systematic uncertainty from the trigger effect was quadratically summed from two sources. The first source is the $E_t$  and Veto trigger, which requires the energy deposit in CSI to be larger than 550 MeV and no on-time hits in CV, NCC, CC03--6, MB or IB detectors respectively. The second source is the clustering trigger. These triggers may accidentally reject events that would normally be accepted in the offline analysis, and this effect was studied using minimum bias data. This data was taken by requiring at least 400 MeV energy deposit in CSI, and no veto or clustering requirements. The relative difference in acceptance in minimum bias data and after applying the standard trigger condition is quoted as the systematic uncertainty (4.1\%). Furthermore, additional systematics were considered in the calculation of the signal acceptance ($A_{sig}$) since additional selection criteria had been used. The systematic uncertainties were evaluated using $\ppp$ data and Monte Carlo as in the $K_L$ yield calculation. The systematic uncertainty in $A_{sig}$ was dominated by $\chi^2_{fit}(3\pi^0)$, and was between 5--8.8\%. Other systematics in $A_{sig}$ include the relative difference in exclusive acceptance of data and Monte Carlo for the FPSD (1.7\%), CSDDL (0.1\%), and $\chi^2_{fit}(2\pi^0\gamma\gamma)$ (1.6\%). The systematic uncertainty labeled ``Quality Cuts'' represents the quadratic sum of uncertainties associated with $\chi^2_{fit}(2\pi^0\gamma\gamma)$, FPSD, CSDDL, and the $\chi^2$ selections used to obtain the $K_L$ flux. The total systematic uncertainty in the SES was 8.9--11.5\% after quadratically summing all of the sources. The SES varies as a function of $X$ mass, with the details shown in Fig.~\ref{fig:SES}.

\textit{Results---}
The background expectation was dominated by $\ppp$ and $\ppgg$ decays, and the level expected for each analysis is shown in Table~\ref{tab:results}. The systematic uncertainty for $N_{bkg}$ is given by the uncertainty in the scale factor $F$, which was between 9--23\%. The results were obtained by opening the signal regions defined by each $X$ mass ROI, while for the $\ppgg$ analysis the entire high mass region was opened at once. Depending on the analysis, the $\chi^2_{fit}(2\pi^0\gamma\gamma) \in [3,9]$, $\chi^2_{fit}(3\pi^0)\in[115,188]$, and the CSDDL threshold was between 0.5--0.9. The results are summarized in Table~\ref{tab:results}. 
\begin{table}[h]
\caption{95\% C.L. upper limits on the branching ratio for each analysis, the number of events observed in data, the expected number of $\ppp$ background, and the expected number of $\ppgg$. The mass, $M_X$, is in units of MeV/$c^2$.}
\def\arraystretch{1.3}
\begin{center}
\scalebox{0.73}{
\begin{tabular}{|| c | c | c | c | c ||} 
 \hline 
    Analysis & $N_{obs}$ & $N_{\ppp}$ & $N_{\ppgg}$ & BR U.L. $(\times 10^{-7})$ \\ [0.5ex] 
 \hline\hline
 $M_X = 160$ & 0 & $<0.34$ (90\% C.L.) & $0.016 \pm 0.003$ & $  < 20.1$ \\ 
\hline
$M_X = 165$ & 0 & $0.41 \pm 0.22$ & $0.09 \pm 0.01$ & $  < 2.95$ \\ 
\hline
$M_X = 170$ & 0 & $0.93 \pm 0.35$ & $0.13 \pm 0.01$ & $  < 1.78$ \\ 
\hline
$M_X = 175$ & 1 & $0.85 \pm 0.35$ & $0.13 \pm 0.01$ & $  < 2.31$ \\ 
\hline
$M_X = 180$ & 2 & $0.19 \pm 0.14$ & $0.11 \pm 0.01$ & $  < 2.82$ \\ 
\hline
$M_X = 185$ & 0 & $<0.28$ (90\% C.L.) & $0.056 \pm 0.008$ & $  < 1.27$ \\ 
\hline
$M_X = 190$ & 0 & $<0.38$ (90\% C.L.) & $0.036 \pm 0.006$ & $  < 1.42$ \\ 
\hline
$M_X = 195$ & 0 & $<0.39$ (90\% C.L.) & $0.033 \pm 0.006$ & $  < 1.45$ \\ 
\hline
$M_X = 200$ & 0 & $0.28 \pm 0.19$ & $0.022 \pm 0.004$ & $  < 1.15$ \\ 
\hline
$M_X = 205$ & 0 & $0.46 \pm 0.24$ & $0.019 \pm 0.004$ & $  < 1.02$ \\ 
\hline
$M_X = 210$ & 1 & $0.19 \pm 0.13$ & $< 0.01$ & $  < 1.25$ \\ 
\hline
$M_X = 215$ & 0 & $0.10 \pm 0.10$ & $< 0.01$ & $  < 0.903$ \\ 
\hline
$M_X = 220$ & 0 & $<0.30$ (90\% C.L.) & $< 0.01$ & $  < 1.24$ 
\\
\hline
\addlinespace[0.2cm]
\hline
 $\ppgg$ & 0 & $1.09 \pm 0.52$ & $0.24 \pm 0.03$ & $< 16.9$ \\
 \hline 
\end{tabular}}
\label{tab:results}
\end{center}
\end{table}

\begin{figure}[h]
\includegraphics[width=0.42\textwidth]{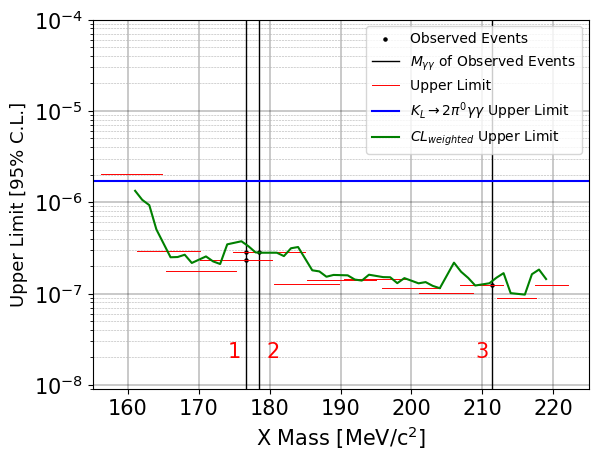}
  \caption{95\% C.L. upper limits. The width of each discrete line shows the ROI for the respective analysis. The points correspond to the three unique observed events and their invariant mass.}
  \label{fig:Upper_limits}
\end{figure}

$N_{obs}$ corresponds to the number of events observed in data within the ROI for that specific analysis. The upper limit was determined using the modified frequentist approach and evaluating $CL_s$~\cite{cls_method}. The number of background events is $N_{\ppp}$ for the analysis of $\ppgg$, and it is the sum of $N_{\ppp}$ and $N_{\ppgg}$ for the analysis of $\ppX$. The quadratic sum of the uncertainty in $N_{bkg}$ and SES are introduced as nuisance parameters. 
Three unique events were observed in the signal regions. The event observed in the analysis of $M_X = 175$ MeV/$c^2$ was also observed in the $M_X=180$ MeV/$c^2$ analysis. This was due to overlapping signal regions and the correlation of selection criteria.\mycomment{ The two events observed in the $M_X=180$ MeV/$c^2$ analysis had an invariant mass of 176.64 and 178.52 MeV/$c^2$ respectively, with the former being the event observed in the 175 MeV/$c^2$ analysis.} To account for the correlation between events and provide a clear upper limit distribution, a weighted $CL_s$ was evaluated. Monte Carlo simulation was performed for all integer $X$ masses between those studied in Table~\ref{tab:results}. The $CL_s$ is evaluated based on the signal acceptance of the integer test mass within each ROI. The 
Each integer $X$ mass is labeled by $m$, and the weighted $CL_s$ is evaluated as,
\begin{equation}
    CL_{weighted}(m) = \frac{\sum_iCL_{s,i}(m)h_i}{\sum_ih_i} \hspace{0.2cm} . 
    \label{eq:weighted_cl_ratio}
\end{equation}
where $h_i$ is the probability density function of the fitted Gaussian at the $i$-th analysis mean. The signal branching ratio assumption which gives a $CL_{weighted} = 5\%$ is the 95\% C.L. upper limit. The results are summarized in Fig.~\ref{fig:Upper_limits}.

%\section{Conclusion and Prospects}
\textit{Conclusion and Prospects---}We have performed a search for $\ppX$ where $X$ decays to $\gamma\gamma$ promptly, and $\ppgg$ at the KOTO experiment using the data taken in 2021. The range of $X$ mass studied was in the range 160--220 MeV/$c^2$, with better sensitivity farther from the pion mass. Three events were observed in the signal regions, with two events near an $X$ mass of 177 MeV/$c^2$. This result led to a range of upper limits on the branching ratio, BR($\ppX$) $< (1\text{--}20) \times 10^{-7}$ at the 95\% C.L. and may directly constrain models where ALP production from neutral three-body decays dominates the two-body modes~\cite{flavourviolation_alp_balkin}. No events were observed for the analysis of $\ppgg$, setting an upper limit on the branching ratio, BR($\ppgg$) $< 1.69 \times 10^{-6}$ at the 95\% C.L. 
\mycomment{The statistics of six cluster data taken by KOTO in 2024--2025 is twice that shown in this analysis, allowing for a deeper search for these modes. These studies are not limited by background contamination, and proportional improvements are expected with more data. This analysis will not be limited by computing power with the aid of the Open Science Grid (OSG) since the time to obtain an adequate simulated $\ppp$ background sample is still reasonable.}  

%\section{Acknowledgements}
%TC:ignore
\textit{Acknowledgements---}We would like to express our gratitude to all members of the J-PARC Accelerator and Hadron Experimental Facility groups for their support. We also thank the KEK Computing Research Center for KEKCC, and the National Institute of Informatics for SINET4. The high throughput simulation was performed with the aid of the University of Chicago Computational Institute and the Open Science Grid Consortium. This material is based upon work supported by the Ministry of Education, Culture, Sports, Science, and Technology (MEXT) of Japan and the Japan Society for the Promotion of Science (JSPS) under KAKENHI Grant Numbers JP16H06343 and JP21H04995 and through the Japan-U.S. Cooperative Research Program in High Energy Physics; the U.S. Department of Energy, Office of Science, Office of High Energy Physics, under Awards No. DE-SC0009798; the National Science and Technology Council (NSTC) and Ministry of Education (MOE) in Taiwan, under Grant Numbers NSTC-108-2112-M-002-001, NSTC-109-2112-M-002-021, NSTC-110-2112-M-002-020, NSTC-111-2112-M-002-032, NSTC-114-2112-M-018-005-MY3, MOE-109L892105, and MOE-110L890205 through National Taiwan University; the National Research Foundation of Korea under Grant Numbers 2020R1A3B2079993, RS-2022-NR070836, and RS-2025-00556834.
%TC:endignore
\appendix

%\nocite{*}
\bibliographystyle{apsrev4-2}
\bibliography{apssamp}% Produces the bibliography via BibTeX.

\end{document}